\documentclass[12pt]{iopart}

\def\Journal#1#2#3#4{{#4} {\it #1} {\bf #2}, #3 }

\begin{document}

\letter{Tidal effects cannot be absent in a vacuum}

\author{Norbert Van den Bergh}

\address{Faculty of Applied Sciences TW16, Gent University, Galglaan 2, 9000 Gent, Belgium}

\begin{abstract}
It is shown that there are no vacuum space-times (with or without
cosmological constant) for which the Weyl-tensor is purely
gravito-magnetic with respect to a congruence of freely falling
observers.

\end{abstract}

\pacs{0420}



\section{Introduction}
Non-conformally flat space-times for which the metric is an exact
solution of the Einstein field equations
\begin{equation}
G_{ab}\equiv R_{ab}-\frac{1}{2}R g_{ab}+\Lambda g_{ab}=T_{ab}
\end{equation}
and in which there exists a family of observers with 4-velocity
$u^a$ ($u_au^a=-1$) such that the gravito-electric (or tidal) part
of the Weyl-tensor vanishes,
\begin{equation}\label{E_ab}
E_{ac}\equiv C_{abcd} u^b u^d=0,
\end{equation}
are called purely gravito-magnetic space-times. They are
remarkable as the remaining gravito-magnetic part of the
Weyl-tensor,
\begin{equation}
H_{ac}\equiv C^*_{abcd}u^b u^d ,
\end{equation}
does not appear in the equation of geodesic deviation, which
implies that in a purely gravito-magnetic vacuum a congruence of
observers would exist for which the geodesic deviation would be
identically zero:
\begin{equation}
\frac{D^2 \mathbf{\xi}}{d \tau^2}\equiv \mathbf{E}.\mathbf{\xi} =
0
\end{equation}
 It has been conjectured that purely gravito-magnetic vacuum
space-times simply do not exist~\cite{McIntosh, Maartens}, but so
far a complete proof has not been given. A partial proof exists
for the special cases where the Petrov type is D~\cite{McIntosh},
or where the timelike congruence $\mathbf{u}$ is
shear-free~\cite{Haddow} or normal~\cite{VdB}. The latter results
extend earlier work on normal and shear-free congruences
~\cite{Trumper} and were generalised recently also to space-times
in which there are less stringent restrictions on the shear and
vorticity tensors~\cite{Ferrando2}. A clear indication that the
field equations for a purely gravito-magnetic vacuum probably are
not consistent when the congruence is geodesic, was given in
\cite{Maartens}, where it was shown that for a dust filled
universe (hence $\dot{\mathbf{u}}=0$) a complicated chain of
integrability conditions has to be satisfied: although the
analysis was done for vanishing vorticity only, the reasoning
suggested a possible way of attack. It is the purpose of the
present paper to demonstrate explicitly that the equations are
indeed inconsistent, at least for vacuum, with or without
cosmological constant.

\section{Relevant equations}

I present below the relevant dynamical equations for a purely
gravito-magnetic vacuum space-time in which the timelike
congruence $u^a$ is geodesic. As in \cite{VdB} I will follow the
notations and conventions of the orthonormal tetrad
formalism~\cite{MacCallum}, with the coefficients $n_{aa}$ being redefined as follows:
\begin{equation}
n_{11}=(n_2+n_3)/2,\ n_{22}=(n_3+n_1)/2, \ n_{33}=(n_1+n_2)/2
\end{equation}
and with the tetrad being specified as an eigenframe of $H_{ab}$.
The system of equations being SO(3)-invariant, each triplet of
equations will be represented by a single equation. The vanishing
of the gravito-electric part of the Weyl-tensor can then be
expressed by the 9 equations
\begin{eqnarray}
\fl E_{11}\equiv -\partial_0
\theta_1-\theta_1^2-\sigma_{12}^2-\sigma_{13}^2+\omega_2^2+\omega_3^2 + 2 \sigma_{12}\Omega_3
-2\sigma_{13}\Omega_2+\frac{1}{3} \Lambda =0 \\
\fl E_{12}\equiv -\partial_0
(\sigma_{12}+\omega_3)-(\theta_1+\theta_2)(\sigma_{12}+\omega_3)-(\sigma_{13}-\omega_2)
( \sigma_{23}-\omega_1)\nonumber \\
+\Omega_1 (\sigma_{13}-\omega_2)-\Omega_2 (\sigma_{23}-\omega_1)
+\Omega_3 (\theta_2-\theta_1) =0 \\
\fl E_{21}\equiv -\partial_0
(\sigma_{12}-\omega_3)-(\theta_1+\theta_2)(\sigma_{12}-\omega_3)-(\sigma_{23}+\omega_1)
(\sigma_{13}+\omega_2)
\nonumber \\
+\Omega_1 (\sigma_{13}+\omega_2)-\Omega_2
(\sigma_{23}+\omega_1)+\Omega_3 (\theta_2-\theta_1) =0
\end{eqnarray}
The vanishing of the off-diagonal components of $H_{ab}$ on the other hand leads to
\begin{eqnarray}
\fl H_{12}\equiv -\partial_0(n_{12}+a_3) - \partial_1 \Omega_2 -
\theta_1 (n_{12}+a_3)-(n_{23}-a_1)(\sigma_{13}+\omega_2)+\frac{1}{2}n_2(\sigma_{12}-\omega_3)\nonumber \\
+\Omega_1 (n_{13}-a_2) -\Omega_2 (n_{23}-a_1 )
+\frac{1}{2} \Omega_3 (n_1-n_2) =0 \\
\fl H_{21}\equiv -\partial_0(n_{12}-a_3) - \partial_2 \Omega_1
-\theta_2 (n_{12}-a_3 )
-(n_{13}+a_2)(\sigma_{23}-\omega_1)+\frac{1}{2}n_1(\sigma_{12}+\omega_3)\nonumber \\
+\Omega_1 (n_{13}+a_2) -\Omega_2 (n_{23}+a_1 ) +\frac{1}{2}
\Omega_3 (n_1-n_2) =0
\end{eqnarray}
Together with the Jacobi-identities (which guarantee the symmetry
of $E_{ab}$ and $H_{ab}$), we obtain from these equations the
evolution for $\theta_a$, $\sigma_{ab}$, $\omega_a$, $a_a$,
$n_{ab}$ and $n_a$:
\begin{eqnarray}\label{eq11}
\fl \partial_0 n_1 = 2 \partial_1
(\omega_1+\Omega_1)+2\partial_2\sigma_{13}-2\partial_3\sigma_{12}
+4 (\omega_2+\Omega_2) n_{13}- 4(\omega_3+\Omega_3) n_{12} \nonumber \\
-n_1 \theta_1-n_2(\theta_1-\theta_3)-n_3 (\theta_1-\theta_2)+4
n_{23}\sigma_{23}
\end{eqnarray}
From (\ref{eq11}) one can eliminate the curl of the shear by using
the diagonal components of $H_{ab}$:
\begin{eqnarray}
\fl H_{11} = \partial_2\sigma_{13}-\partial_3
\sigma_{12}+\partial_1\omega_1-\frac{1}{2}(\theta_1
(n_2+n_3)-\theta_2n_3-\theta_3 n_2)+2n_{23}\sigma_{23}\nonumber \\
-n_{12}(\sigma_{12}+\omega_3)-a_2(\sigma_{13}+\omega_2)-n_{13}
(\sigma_{13}-\omega_2)+a_3(\sigma_{12}-\omega_3)
\end{eqnarray}
The remaining Jacobi identities contain spatial gradients of the
kinematical scalars only and can be used to simplify the
integrability conditions which result by considering the following
commutators:
\begin{equation}
[\partial_0,\partial_1] \theta_2-[\partial_0,\partial_2]
(\sigma_{12}-\omega_3) \ \ \textrm{and}\ \ [\partial_0,\partial_1]
\theta_3-[\partial_0,\partial_3] (\sigma_{13}+\omega_2)
\end{equation}
One obtains then three pairs of equations,
\begin{eqnarray}
(\sigma_{23}-\omega_1+\Omega_1)H_{11}-(\sigma_{23}-\omega_1-2\Omega_1)H_{22}
= 0 \\
(2\sigma_{23}+2\omega_1+\Omega_1)H_{11}+(\sigma_{23}+\omega_1+2\Omega_1)H_{22}=0
\end{eqnarray}
in which one recognizes the familiar relation~\cite{Maartens1,
Haddow, Ferrando2} between $\mathbf{\sigma}$, $\mathbf{\omega}$
and $\mathbf{H}$,
\begin{equation}\label{Haddow}
\mathbf{\sigma} \times \mathbf{H} = 3
\mathbf{\omega}\cdot\mathbf{H},
\end{equation}
together with a relation between $\mathbf{\sigma}$,
$\mathbf{\omega}$ and the rotation rate $\mathbf{\Omega}$ of the
$\mathbf{H}$-eigenframe with respect to a Fermi-propagated triad:
\begin{equation}\label{Omega}
\omega_1^2+2 \omega_1 \Omega_1 = \sigma_{23}^2
\end{equation}
(+ cyclic permutations). In fact the latter equations hold also in
the general vacuum case, when the acceleration is non-zero! The
last bit of information we need is the time evolution of the
curvature, in the form
\begin{equation}
\partial_0 H_{11}=\theta_2 (H_{33}-H_{11})+\theta_3(H_{22}-
H_{11}),
\end{equation}
which can be obtained directly by considering the
$[\partial_0,\partial_2](\sigma_{13}-\omega_2)-[\partial_0,\partial_3](\sigma_{12}+\omega_3)$
commutators. Using these one can simplify the equations which
result by substituting the expressions obtained from
(\ref{Haddow}, \ref{Omega}) for $\mathbf{\sigma}$ and
$\mathbf{\Omega}$, namely
\begin{eqnarray}\label{rels}
\sigma_{12}=3\omega_3 \frac{H_{11}+H_{22}}{H_{11}-H_{22}}\\
\Omega_1=2 \omega_1
\frac{(H_{11}-H_{22})(H_{11}-H_{33})}{(H_{22}-H_{33})^2}
\end{eqnarray}
in the evolution equations for the shear. Note that $\mathbf{H}$
is not allowed to have equal eigenvalues~\cite{McIntosh}. One
finds then the following algebraic relations between
$\mathbf{\omega}$ and $\mathbf{H}$:
\begin{eqnarray}\label{tees}
\fl 2\omega_3(H_{22}-H_{33})(H_{11}-H_{33})[
\theta_1(H_{22}-H_{33})+\theta_2(H_{11}-H_{33})+3\theta_3(H_{11}-H_{22})]
\nonumber \\
+3\omega_1 \omega_2 (H_{11}-H_{22})(5
H_{11}^2+8H_{11}H_{22}+5H_{22}^2)=0
\end{eqnarray}
Eliminating $\theta_a$ from the latter equation and its cyclic
permutations results in
\begin{eqnarray}
\fl \left( 2\,{H_{{11}}}^{2}+2\,H_{{11}}H_{{22}}+5\,{H_{{22}}}^{2}
\right)  \left( H_{{11}}-H_{{33}} \right)
^{4}{\omega_{{1}}}^{2}{ \omega_{{3}}}^{2}\nonumber \\
+ \left(
2\,{H_{{22}}}^{2}+2\,H_{{22}}H_{{ 33}}+5\,{H_{{33}}}^{2} \right)
\left( H_{{22}}-H_{{11}} \right) ^{4}{
\omega_{{2}}}^{2}{\omega_{{1}}}^{2} \nonumber \\
+ \left( 2\, {H_{{33}}}^{2}+2\,H_{{11}}H_{{33}}+5\,{H_{{11}}}^{2}
\right) \left( H_{{22}}-H_{{33}} \right) ^{4}
{\omega_{ {3}}}^{2}{\omega_{{2}}}^{2}
=0
\end{eqnarray}
All the coefficients in the above expression are strictly positive
(as a degenerate $\mathbf{H}$ is not allowed), finishing the proof
that a purely magnetic vacuum is
inconsistent with the assumption of a geodesic congruence. \\
Note that the above analysis breaks down when one or two
components of the vorticity vanish. Taking into account however
the evolution equations for the vorticity, it follows that at
least two components must vanish, say $\omega_2 = \omega_3 = 0$,
such that in stead of (\ref{rels}) and its cyclic permutations one
obtains only a single relation (\ref{rels}). Substituting this in
the evolution for the shear leads then to a single algebraic
relation between the $H_{ab}$ and $\theta_a$, a further time
derivative of which is needed to obtain an inconsistency between
the signs of the involved $\omega_1^2$ and curvature terms.

\end{document}